\documentclass[5p,twocolumn,10pt,authoryear]{elsarticle}

\usepackage[utf8]{inputenc} 
\usepackage{amsmath}
\usepackage{amsfonts}
\usepackage{amssymb}

\usepackage{multirow}
\usepackage{xcolor}

\newcommand{\degree}{\ensuremath{^{\circ}}} 
\newcommand{\rmsub}[1]{\ensuremath{_{\mathrm{#1}}}}

\newcommand{\figref}[1]{Fig.~\ref{fig:#1}}

\newcommand{\eqnrefjustpar}[1]{(\ref{eq:#1})}
\newcommand{\eqnref}[1]{Eq.~\eqnrefjustpar{#1}}
\newcommand{\eqnsref}[1]{Eqs.~\eqnrefjustpar{#1}}

\newcommand{\werrhat}{\ensuremath{\hat{\omega}\rmsub{err}}}
\newcommand{\Wvis}{\ensuremath{W\rmsub{vis}}}
\newcommand{\Wves}{\ensuremath{W\rmsub{ves}}}
\newcommand{\wvishat}{\ensuremath{\hat{\omega}\rmsub{vis}}}
\newcommand{\wveshat}{\ensuremath{\hat{\omega}\rmsub{ves}}}
\newcommand{\wvis}{\ensuremath{\omega\rmsub{vis}}}
\newcommand{\wves}{\ensuremath{\omega\rmsub{ves}}}
\newcommand{\wbody}{\ensuremath{\omega\rmsub{body}}}
\newcommand{\nuvis}{\ensuremath{\nu\rmsub{vis}}}
\newcommand{\nuves}{\ensuremath{\nu\rmsub{ves}}}
\newcommand{\sigmavis}{\ensuremath{\sigma\rmsub{vis}}}
\newcommand{\sigmaves}{\ensuremath{\sigma\rmsub{ves}}}
\newcommand{\rvis}{\ensuremath{r\rmsub{vis}}}
\newcommand{\rves}{\ensuremath{r\rmsub{ves}}}
\newcommand{\Jtot}{\ensuremath{J\rmsub{tot}}}
\newcommand{\Jpath}{\ensuremath{J\rmsub{path}}}
\newcommand{\Jsteer}{\ensuremath{J\rmsub{steer}}}
\newcommand{\kpath}{\ensuremath{k\rmsub{path}}}
\newcommand{\ksteer}{\ensuremath{k\rmsub{steer}}}
\newcommand{\TP}{\ensuremath{T\rmsub{P}}}

\newcommand{\inquotes}[1]{\lq\lq #1\rq\rq}

\begin{document}

\begin{frontmatter}

\title{Modelling visual-vestibular integration and behavioural adaptation in the driving simulator}

\author[1]{Gustav Markkula}
\author[1]{Richard Romano}
\author[2]{Rachel Waldram}
\author[1]{Oscar Giles}
\author[2]{Callum Mole}
\author[2]{Richard Wilkie}
\address[1]{Institute for Transport Studies, University of Leeds, LS2 9JT, Leeds, United Kingdom}
\address[2]{School of Psychology, University of Leeds, LS2 9JT, Leeds, United Kingdom}

\begin{abstract}
It is well established that not only vision but also other sensory modalities affect drivers’ control of their vehicles, and that drivers adapt over time to persistent changes in sensory cues (for example in driving simulators), but the mechanisms underlying these behavioural phenomena are poorly understood. Here, we consider the existing literature on how driver steering in slalom tasks is affected by the down-scaling of vestibular cues, and propose a driver model that can explain the empirically observed effects, namely: decreased task performance and increased steering effort during initial exposure, followed by a partial reversal of these effects as task exposure is prolonged. Unexpectedly, the model also reproduced another empirical finding: a local optimum for motion down-scaling, where path-tracking is better than when one-to-one motion cues are available. Overall, the results imply that: (1) drivers make direct use of vestibular information as part of determining appropriate steering, and (2) motion down-scaling causes a yaw rate underestimation phenomenon, where drivers behave as if the simulated vehicle is rotating more slowly than it is. However, (3) in the slalom task, a certain degree of such yaw rate underestimation is beneficial to path tracking performance. Furthermore, (4) behavioural adaptation, as empirically observed in slalom tasks, may occur due to (a) down-weighting of vestibular cues, and/or (b) increased sensitivity to control errors, in determining when to adjust steering and by how much, but (c) seemingly not in the form of a full compensatory rescaling of the received vestibular input. The analyses presented here provide new insights and hypotheses about simulator driving, and the developed models can be used to support research on multisensory integration and behavioural adaptation in both driving and other task domains.  
\end{abstract}


\end{frontmatter}


\section{Introduction}

Driving simulators can be valuable tools for research on driver behaviour, industrial prototyping of vehicles, and training of drivers \citep{FisherEtAl2011}, but only as long as the realism of the simulated driving is satisfactory for the application at hand. For this reason, research into driving simulator realism and validity is an active field of work. This is true not least when it comes to the \emph{motion cueing} in motion-based simulators; i.e., how to best move the simulator within its limited motion envelope, to nevertheless create a maximally realistic experience of vehicle movement for the driver, and to elicit objective driver behaviour that is successful, and ideally similar to that in a real vehicle \citep{SieglerEtAl2001, FischerEtAl2016, SalisburyAndLimebeer2017}. 

Typical motion cueing algorithms attempt to leverage the properties and limitations of the vestibular (motion) sensory organs in the inner ear, the \emph{otoliths} and \emph{semicircular canals}, and these systems are relatively well understood and have been modelled in mathematical detail \citep{Hosman1996, NashEtAl2016}. However, much less is known about (i) how drivers then integrate this vestibular information with information from other sensory modalities (e.g., vision) to support vehicle control, (ii) how these processes are affected by the specific nature of the non-perfect motion cues being provided, and (iii) how drivers adapt over time to such imperfections. 
A quantitative model of driver behaviour that successfully captured such aspects of multisensory integration and behavioural adaptation would be an eminent tool for improving simulators and motion cueing algorithms.
The present paper proposes such a model, and tests it against empirical findings from the literature on driving in slalom tasks. Below, the specific aims and structure of this paper will be described, after first providing brief overviews of existing empirical knowledge about drivers' response to down-scaled motion cues (a special case of motion cueing), and existing models of multisensory integration and driver steering.

\subsection{Studies on simulator motion scaling}

Most motion cueing algorithms include some element of \emph{linear down-scaling} of the actual motion of the vehicle, to stay within the motion envelope of the simulator, even all the way down to zero motion scaling in fixed base simulators. One often observed effect of down-scaled or zero motion is that drivers adopt more aggressive driving strategies \citep[higher speeds and/or tighter curve-taking; ][]{SieglerEtAl2001, Jamson2010, CorreiaGracioEtAl2011, BerthozEtAl2013}, and that the control of the vehicle generally becomes less accurate and more effortful \citep{RepaEtAl1982, Jamson2010, BerthozEtAl2013}. This may however be dependent on the specific driving task, since \citet{WilkieAndWann2005} found no effect of motion scaling on behaviour in a simple task where participants steered towards a single target.

One type of task where reliable effects have been found however, and which is of particular interest in the context of motion scaling is \emph{slalom} driving. This task is experimentally useful because, when driven at constant or near-constant speed in a simulator, the motion cueing algorithm can be simplified to direct linear scaling only, isolating the effect of motion scaling on behaviour from other types of motion manipulations that are otherwise often applied \citep[e.g., \emph{tilt coordination} and \emph{washout}; ][]{Jamson2010}. Consequently, many authors have studied how driver behaviour in slalom tasks is affected by variations in motion scaling, providing converging evidence for a set of behavioural phenomena: Steering effort, for example measured as steering reversal rates or high frequency steering content, generally increases when motion cues are removed \citep{FeenstraEtAl2010, CorreiaGracioEtAl2011, SavonaEtAl2014b}, and task performance, objectively measured or subjectively assessed, generally deteriorates \citep{CorreiaGracioEtAl2011, BerthozEtAl2013}, although not always \citep{SavonaEtAl2014b}. Taken together, these findings suggest that in normal driving, drivers integrate visual and vestibular cues, and when vestibular cues are removed, this makes task performance more challenging. However, after repeated exposure to the slalom task, control efforts decrease \citep{FeenstraEtAl2010} and task performance improves \citep{CorreiaGracioEtAl2011}, suggesting some form of behavioural adaptation over time to the lack of vestibular cues. For motion scaling between zero and one, there are reports from several studies of a local optimum, in the 0.4-0.8 range, where task performance and subjective preferences peak \citep{BerthozEtAl2013, SavonaEtAl2014b}, and in one case this local optimum was also observed for steering effort \citep{SavonaEtAl2014b}. Different theories have been proposed for why drivers prefer and perform the slalom best at somewhat down-scaled motion cues; here a novel explanation of this phenomenon will be provided, suggesting that it may be a direct effect of multisensory integration.

\subsection{Models of multisensory integration and steering}

\begin{figure}
	\centering
	\includegraphics[width=\columnwidth]{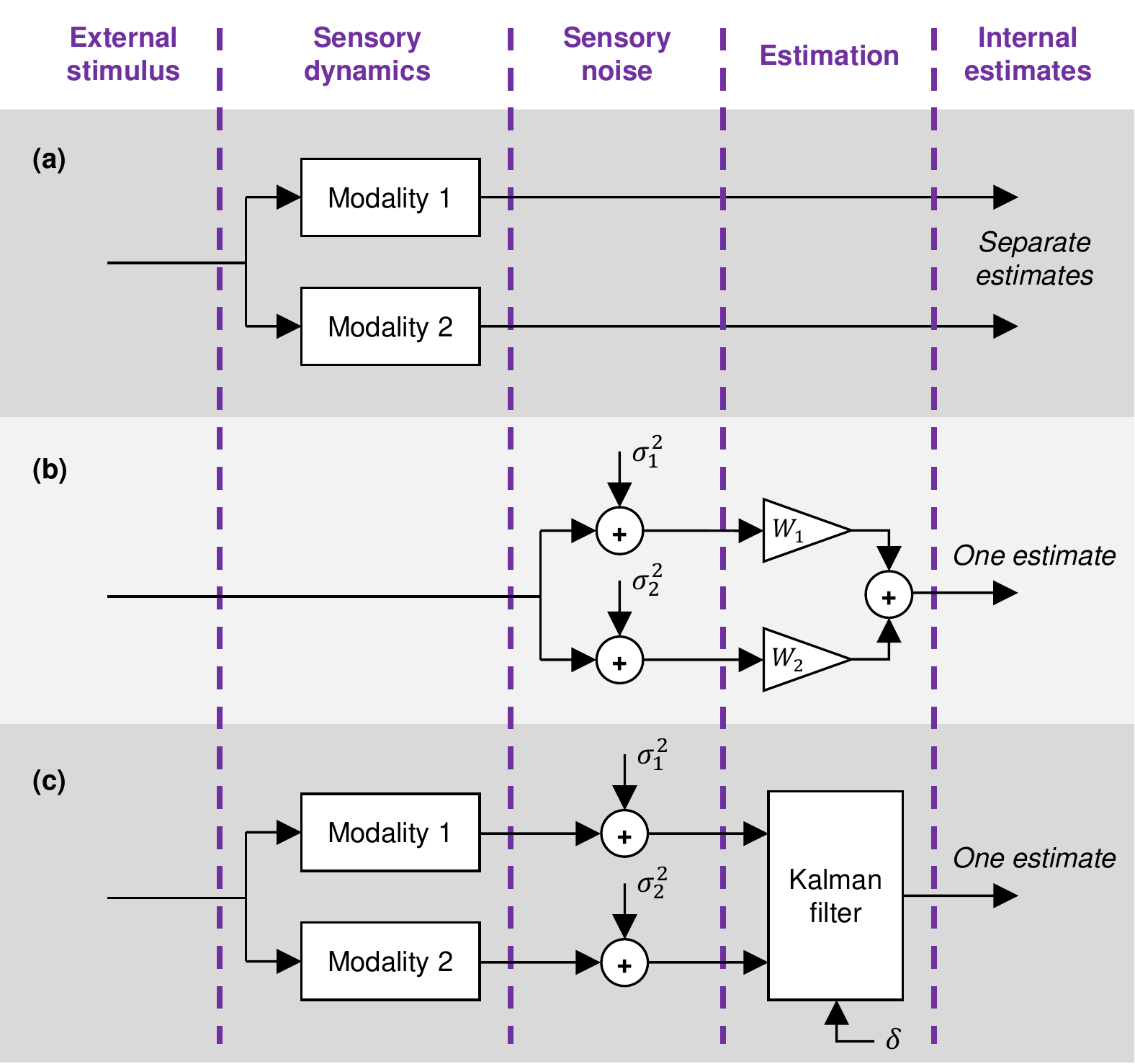}
	\caption{Three different types of modelling schemes for multisensory integration. (a) Sensory dynamics models only; common in aircraft pilot models in research on flight simulators. (b) Modality-dependent sensory noise (variances $\sigma_1^2$ and $\sigma_2^2$) and optimal cue integration; common in psychophysical studies of multisensory integration. (c) Sensory dynamics models, modality-dependent sensory noise, and optimal state estimation, including both optimal cue integration and model-based predictions based on own past behaviour ($\delta$); this scheme is common in optimal control theory models of multimodal sensorimotor control.}
	\label{fig:MultisensoryIntegrationSchemes}
\end{figure}

There is a large literature on driver steering models \citep[see, e.g., the reviews by ][]{PlochlAndEdelmann2007, SteenEtAl2011, Lappi2014}. A recurring theme in this literature has been one inspired from ecological psychology, the study of what exact visual quantities are being sensed and used by drivers \citep{Lappi2014}, and there has also been a recent surge in interest in modelling steering as \emph{intermittent} control \citep[e.g., ][]{GordonAndZhang2015, JohnsAndCole2015, MarkkulaEtAl2018}. However, steering models have typically not considered visual-vestibular sensory integration. In contrast, in the aviation domain, where flight simulators play a very important role in pilot training and certification, there have been substantial efforts to develop multisensory models \citep[e.g., ][]{Hosman1996, ZaalEtAl2009, MulderEtAl2013}. In these pilot models, there has been a strong emphasis on \emph{sensory dynamics models}, especially for the vestibular senses. These sensory dynamics models aim to capture how the neural firing rates along afferents leaving the inner ear are transformed compared to the body accelerations and rotations being experienced by the organism \citep[][provide a review]{NashEtAl2016}. In this type of approach to multisensory modelling, illustrated in \figref{MultisensoryIntegrationSchemes}(a), the resulting modality-specific internal estimates are then typically combined in linear control laws to model pilot control actions.

In research focusing on perception and psychophysics, on the other hand, the type of model illustrated in \figref{MultisensoryIntegrationSchemes}(b) has been more common. This model disregards sensory dynamics and instead emphasises the \emph{sensory noise} present in the different modalities; it has been repeatedly shown that humans and other animals often behave as (near-)ideal Bayesian observers, by engaging in (near-)\emph{optimal cue integration}, where the sensory cues are weighted by their reliabilities, i.e., the inverse of their respective noise variances ($W_i \propto 1/\sigma_i^2$ in \figref{MultisensoryIntegrationSchemes}). \citet{FetschEtAl2013} and \citet{NashEtAl2016} provide reviews of this literature from neurobiological and vehicle control perspectives, respectively.

The model illustrated in \figref{MultisensoryIntegrationSchemes}(c) is a combination of the models in (a) and (b), with the optimal cue integration being carried out as part of a Kalman filter. This type of scheme has been common in optimal control theoretic models of various sensorimotor control tasks \citep{vanDerKooijEtAl1999, vanBeersEtAl2002, FranklinAndWolpert2011}, and was recently applied by \citet{NashAndCole2018} in their multisensory optimal control model of driver steering. In these models, the Kalman filter makes use of knowledge of the own past control actions as well as internal forward models of the system being controlled, including the own sensory dynamics. Therefore, as long as any changes in a vehicle's state are mainly caused by the driver's own control, and not by external perturbations such as potholes or wind gusts, the Kalman filter in \figref{MultisensoryIntegrationSchemes}(c) can compensate for the sensory dynamics, essentially reducing the model to the situation in \figref{MultisensoryIntegrationSchemes}(b). Since the empirical literature on slalom driving does not emphasise external perturbations, the scheme in \figref{MultisensoryIntegrationSchemes}(b) will be adopted here.

Predictive forward models, or \emph{generative} models, are also emphasised in so-called \emph{predictive processing} accounts of brain function \citep[e.g., ][]{Friston2005, Bogacz2017}. These accounts suggest neurobiologically plausible mechanisms for how the brain can learn its own sensory reliabilities, by first learning a generative model, and then observing deviations between received and predicted sensory data. This type of mechanism will also be leveraged here.

\subsection{Aims of this paper}

This paper proposes, for the first time, a driver steering model that reproduces, and therefore provides stringent and concrete explanations for, the observed empirical effects of motion down-scaling in simulators, specifically (i) how humans first react to down-scaled motion cues, and (ii) how this behaviour changes with repeated task exposure. To address this aim, the model needs to account both for multisensory integration and some form of behavioural adaptation. Here, the neurobiologically plausible framework for intermittent control proposed by \citet{MarkkulaEtAl2018} was used as a starting point, since it lends itself naturally to a range of plausible potential mechanisms for behavioural adaptation. From a modelling perspective, the novelty lies in the extension of the existing framework with these adaptation mechanisms (as far as we are aware this is the first ever model of behavioural adaptation in steering), as well as with the optimal cue integration mechanisms reviewed above. It was not an original aim of this work to study the phenomenon of a local optimum in slalom motion downscaling, but as will become clear below, the model turns out to be applicable also to this phenomenon.

Below, the steering model is first introduced. Then, results from model simulations are presented, before a discussion and conclusions are provided.

\section{Theory and methods}

\begin{figure*}[t]
	\centering
	\includegraphics[width=\textwidth]{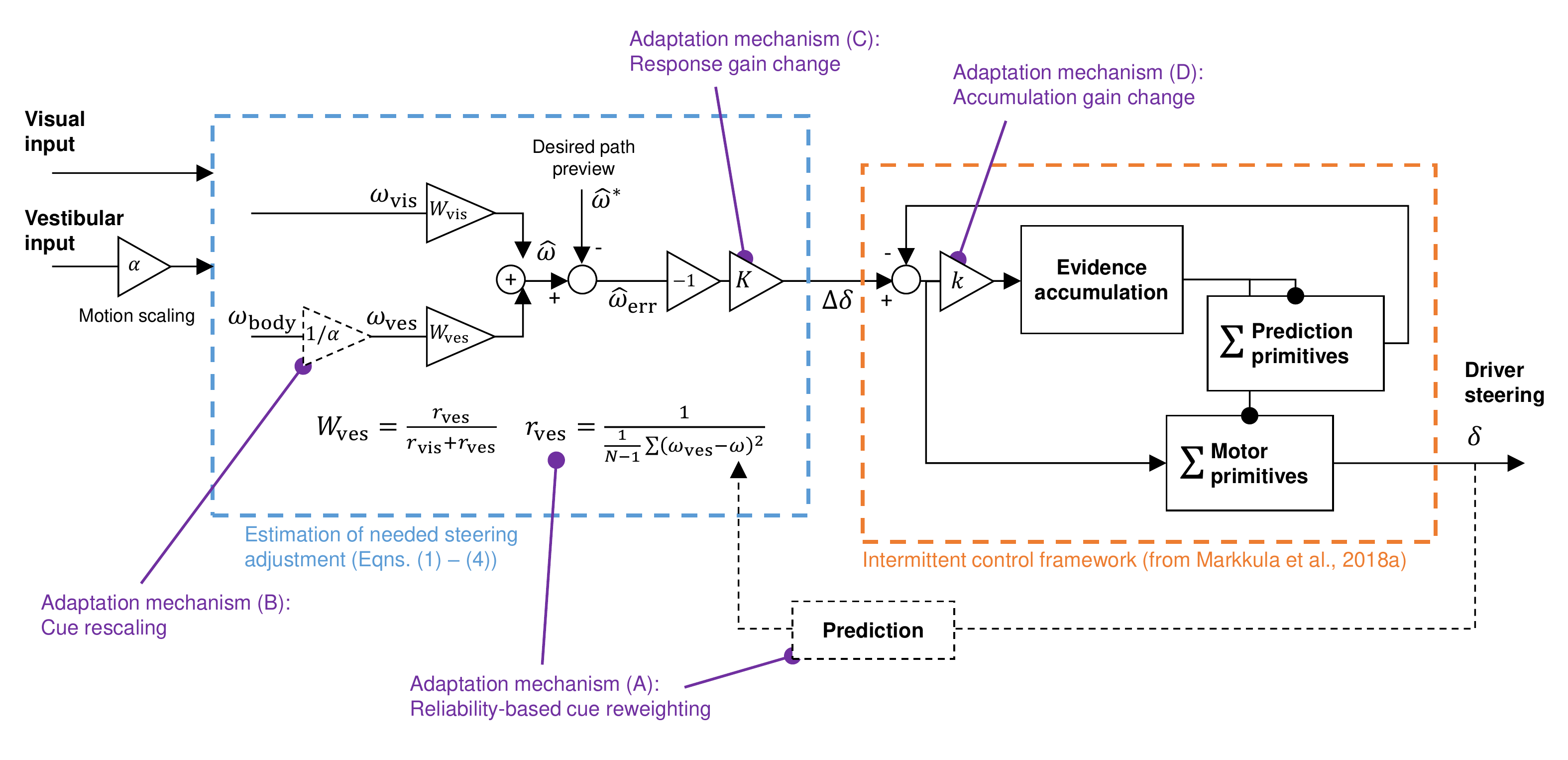}
	\caption{Schematic illustration of the driver steering model tested in this paper. The leftmost part of the figure shows how the model responds to visual and (possibly down-scaled) vestibular cues, creating an estimate of needed steering adjustment $\Delta\delta$ by weighting the modality-specific estimates of yaw rate $\omega$ according to an optimal cue integration scheme (equations shown only for vestibular weights and reliabilities $\Wves$ and $\rves$), and then comparing the integrated estimate to a desired yaw rate $\hat{\omega}^*$. The rightmost part of the figure is a simplified illustration of how the intermittent control framework (described in full detail by \citealp{MarkkulaEtAl2018}) transforms $\Delta\delta$ into the steering wheel angle $\delta$. Four of the tested behavioural adaptation mechanisms are also indicated in the figure. The $1/\alpha$ cue rescaling gain is shown in dashed line because it is only present in the model when that adaptation mechanism is enabled, when it is not, $\wves = \wbody$. The prediction from $\delta$ to $\omega$ is shown in dashed line because it is assumed in the model without being modelled explicitly; i.e., the prediction-based cue reweighting is using the true yaw rate $\omega$ directly.}
	\label{fig:ModelOverview}
\end{figure*}

This section describes the proposed driver steering model, illustrated in \figref{ModelOverview}, as well as the model simulations carried out to study the model's behaviour.

\subsection{Driver steering model}

\subsubsection{Slalom desired path}

The model adopts the commonly used concept of a \emph{desired path} \citep{PlochlAndEdelmann2007} to define the sinusoidal slalom task. The exact setup of this task was here based on \citep{FeenstraEtAl2010}: 62.5 m spacing between cones, 3 m lateral amplitude, and the task was carried out at a constant longitudinal speed of 70 km/h.

\subsubsection{Intermittent control framework}

As mentioned, the model builds on a framework for intermittent sensorimotor control, introduced and described in detail by \citet[][see the rightmost part of \figref{ModelOverview}]{MarkkulaEtAl2018}. In brief, this framework assumes that a continuously calculated estimate of currently needed control adjustment $\Delta \delta$ is compared to a prediction of $\Delta \delta$, to yield a prediction error. This prediction error is then fed, with a gain $k$, into an \emph{evidence accumulation} (also known as \emph{drift diffusion}) step where it is integrated over time to a threshold of $\pm 1$, to decide on when a control adjustment is needed. This is in line with neurobiologically proven mechanisms for how the brain makes decisions based on noisy sensory data \citep{GoldAndShadlen2007, PurcellEtAl2010, RatcliffEtAl2016}.  In the \citet{MarkkulaEtAl2018} framework, when the evidence accumulation has integrated to threshold, the integrator is reset to zero, a control adjustment is initiated, and a prediction is made of how $\Delta \delta$ will be affected over time by the new control adjustment. The adjustment as such is applied in the form of a kinematic \emph{motor primitive} \citep{Giszter2015}, here a fixed-duration ($\Delta T$ = 0.4 s, based on the findings by \citealp{BenderiusAndMarkkula2014}) stepwise change of the steering wheel angle, with a bell-shaped rate profile and with amplitude obtained directly from the current $\Delta \delta$ prediction error, plus signal-dependent motor noise \citep{FranklinAndWolpert2011}. Also the $\Delta \delta$ prediction is obtained as a superposition of stereotyped \inquotes{prediction primitives}, mimicking neurobiologically observed \emph{corollary discharge} \citep{CrapseAndSommer2008, RequarthAndSawtell2014}.

\subsubsection{Needed steering adjustment}

In the engineering literature, there are several driver steering models \citep{GordonAndMagnuski2006, TanAndHuang2012, MarkkulaEtAl2014} on the general form:
\begin{equation}
	\label{eq:DPYRE}
	\Delta\delta = -K \werrhat = -K (\hat{\omega} - \hat{\omega}^*),
\end{equation}
where $\hat{\omega}^*$ and $\hat{\omega}$ are desired and actual vehicle yaw rate, as perceived by the driver, and $K$ is a steering response gain. As mentioned in the Introduction, there is a considerable literature in the ecological psychology tradition, arguing for the importance of considering what types of information is actually perceptually available to the driver, rather than assuming access to \inquotes{engineering quantities} like yaw rate. Interestingly however, it can be shown that models in the ecological psychological literature, emphasising more plausible visual inputs such as sight point rotations, can be rewritten on the form of \eqnref{DPYRE} \citep{Markkula2013}, so it can be argued that this type of control law is perceptually plausible.

Here, the desired yaw rate $\hat{\omega}^*$ is defined as the yaw rate that would take the vehicle back to the desired path in a preview time $\TP$. This formulation has been shown to successfully replicate human slalom steering, with $\TP \in [1.4, 2.2]$ s \citep{MarkkulaEtAl2018THMS}; here $\TP$ = 1.8 s was used. 


\subsubsection{Visual-vestibular integration}

As mentioned above, here we follow the psychophysical literature on optimal cue integration, modelling multisensory integration as operating directly on estimates of the external stimulus, in this case vehicle yaw rate:
\begin{align}
	\label{eq:CueIntegration}
	\hat{\omega} 	& = \Wvis \wvishat + \Wves \wveshat \nonumber \\
								& = \Wvis(\wvis + \nuvis) + \Wves(\wves + \nuves),
\end{align}
where $\wvis = \omega$ and $\wves = \wbody = \alpha \omega$, with $\wbody$ the actual rotation of the driver's body, $\alpha$ the motion scaling being applied, $\omega$ the yaw rate of the simulated vehicle, and $\nuvis$ and $\nuves$ being Gaussian white noise with standard deviations $\sigmavis$ and $\sigmaves$. Optimal cue integration theory prescribes:
\begin{equation}
	\label{eq:OptimalCueIntegration}
	\Wvis = \frac{\rvis}{\rvis + \rves}, \;\; \Wves = \frac{\rves}{\rvis + \rves},
\end{equation}
with $\rvis$ and $\rves$ being the respective sensory reliabilities. Note that here, the otoliths and semicircular canals are in effect subsumed into a single estimate of yaw rate, since this is all the control law in \eqnref{DPYRE} needs (but it may also be noted that in a vehicle, lateral acceleration, as sensed by the otoliths, typically provides good information on yaw rate also).

Thus, overall, what is suggested here (in the leftmost dashed box in \figref{ModelOverview}) is that drivers might behave as if (i) they transform the neural output from their sensory organs into estimates of yaw rates, presumably considering also predictions based on knowledge of past steering input, (ii) integrate these yaw rate estimates as per \eqnsref{CueIntegration} and \eqnrefjustpar{OptimalCueIntegration}, and then (iii) compare the result to a visually estimated desired yaw rate, to calculate the needed steering adjustment as per \eqnref{DPYRE}. Note the, \inquotes{as if} above; it is not assumed that drivers' brains necessarily directly perceive and encode things like desired paths or desired or actual yaw rates, only that they behave as if they do.

Here, $\sigmavis = 0.5$\degree/s was used, loosely based on \cite{NestiEtAl2015}, who found that humans could discriminate between visual rotation stimuli of about 5 \degree/s, at which yaw rates typically peak in the present slalom task, if they were different by 1 \degree/s. The 0.5 \degree/s noise level gives 75 \% correct direction classification, by a drift diffusion model such as in the intermittent control framework used here, of a 1 \degree/s stimuli if presented during 10 s, a similar duration of presentation as in \cite{NestiEtAl2015}. 

As for the vestibular noise, the literature on how visual and vestibular sensory systems compare in terms of delays or noise levels provide conflicting information for different types of experimental condition \cite{NashEtAl2016}, so for simplicity we here assume $\sigmavis = \sigmaves$, to begin with. However, later in this paper we also examine the model's sensitivity to variations in these noise levels. 

It is assumed that when first entering the simulator, the driver's sensory weightings are preset, based on prior driving experience, in line with the true sensory reliabilities ($\rvis = \sigmavis^{-2}$; $\rves = \sigmaves^{-2}$), which with $\sigmavis = \sigmaves$ gives $\Wvis = \Wves = 0.5$.

\subsubsection{Behavioural adaptation}

As also illustrated in \figref{ModelOverview}, a number of mechanisms for behavioural adaptation are assumed to be operating, alone or in combination:

(A) Reliability-based cue reweighting. In line with the predictive processing models mentioned above, this adaptation mechanism assumes that the driver has a generative model that can predict current yaw rate $\omega$ based on past steering input $\delta$, but this prediction is not modelled explicitly here; the model instead operates on true $\omega$ directly. After each simulation run of the slalom, the perceived visual and vestibular yaw rate estimates $\wvis$ and $\wves$ are compared to the predicted (i.e., true) $\omega$, across all $N$ discrete simulation time steps, and the sensory reliabilities are estimated as
\begin{eqnarray}
	\label{eq:PredictionBasedSensoryReliability}
	\rvis = \frac{1}{\frac{1}{N-1}\sum_{i=1}^{N}{(\omega_{\mathrm{vis}, i} - \omega_i)^2}} \nonumber \\ 
	\rves = \frac{1}{\frac{1}{N-1}\sum_{i=1}^{N}{(\omega_{\mathrm{ves}, i} - \omega_i)^2}},
\end{eqnarray}
and the weights $\Wvis$ and $\Wves$ are updated in accordance with \eqnref{OptimalCueIntegration} before the next simulation run. As long as there is no motion down-scaling (i.e., $\alpha = 1$), this will just result in the model retrieving approximate estimates of the true inverse noise variances ($\rvis \approx \sigmavis^{-2}$; $\rves \approx \sigmaves^{-2}$). However, when $\alpha \neq 1$ and $\wves = \alpha \omega$, the estimation of $\rves$ is affected by the resulting persistent bias of $\wves$ away from $\omega$, causing lower vestibular reliabilities $\rves$ the further $\alpha$ diverges from unity, i.e., the further the perceived vestibular cues are from what the driver's brain would expect based on its past experiences from real driving.

(B) Reinterpreting the downscaled vestibular cues, by \emph{rescaling} the mapping from vestibular organ output to $\wves$, i.e., $\wves = \wbody / \alpha$. Since this type of sensory relearning simultaneously scales up the vestibular noise by $1/\alpha$, it is assumed that it would include appropriate reweighting of cues, i.e., mechanism (C) was also included here.

(C) Adapting the steering response gain $K$, i.e., increasing or decreasing how much one changes the steering angle for a given perceived yaw rate error \werrhat. 

(D) Adapting the gain $k$ in the evidence accumulation. This can be thought of as adapting \emph{effort} or \emph{arousal}; increases in $k$ cause the model to apply more frequent (and hence typically smaller) steering adjustments, and vice versa.

The mechanisms (C) and (D) are assumed to operate so as to optimise behaviour towards some goal, here implemented in practice as an exhaustive search over a grid of the optimised parameters, minimising the cost function:
\begin{align}
\label{eq:CostFunction}
	\Jtot 	& = \kpath\Jpath + \ksteer\Jsteer \nonumber \\ 
					& = \kpath\frac{1}{N} \sum_{k=1}^N (y_k - y_k^*)^2 + \ksteer\frac{1}{T}\sum_{i=1}^n g_i^2
\end{align}
for a simulation of duration $T$, with $N$ discrete samples, in which $n$ discrete steering adjustments with amplitudes $g_i$ are applied, and with lateral positions of vehicle and desired path $y_k$ and $y_k^*$, respectively. This is a common form of cost function for optimal control theory models of steering \citep{PlochlAndEdelmann2007, NashAndCole2018}. The weighting parameters were set to $\kpath = 1$ and $\ksteer = 10$ to get similar magnitudes across both terms in typical simulations with the model. Note that it is not implied here that the brain performs grid search to change its behaviour; only that it implements some optimisation mechanism (unspecified here, but see the Discussion) of which the effect is gradual reduction of something like $\Jtot$.

As will be described under Results below, all of the the mechanisms (A), (C), and (D) were found promising individually, but not (B). Therefore, also the following was tested:

(E) All three adaptations (A), (C), and (D) operating simultaneously.

Finally, also this mechanism was tested:

(F) Adapting sensory weights not based estimated reliability (as in mechanism (A)), but instead to minimise $\Jtot$ using the grid search optimisation described above. The purpose of this test was to investigate whether or not the reliability-based cue reweighting (A) yielded similar results as a formal optimisation of the cue weights.

\subsection{Driver-vehicle model simulations}

Simulations were run with a linear vehicle model, fitted to multibody simulations of a Jaguar XF driving slaloms, mostly in the linear tyre regime.

For the gains $k$ and $K$, initial, before-adaptation values for the model were selected as those gains that minimised $\Jtot$ for driving with full motion ($\alpha = 1$); i.e., it was assumed that the driver comes to the simulator with these gains preset based on prior driving experience (and can rapidly adapt to the steering gain of the specific simulated vehicle). In practice, an exhaustive grid of values for both gains was searched, with ten repetitions per gain combination (since the driver model is stochastic) of the $\alpha = 1$ slalom. Optimal model performance was obtained for $k = 300$ (arbitrary units) and $K = 2.04$ s. It may be noted that this latter figure is close to the theoretically optimal steering response gain 1/0.510 s = 1.96 s, with 0.510 s$^{-1}$ being the steady state yaw rate response gain of the vehicle model used in the simulations. 

For motion gains $\alpha \in \{0, 0.2, ..., 1\}$, simulations were then first run without any behavioural adaptations, i.e., without any changes to the model parameters. These simulations served to emulate non-adapted driver behaviour during first exposure to down-scaled motion cues. Then, the the hypothesised behavioural adaptations (A)-(F) described above were applied, and new simulations were run, to study the impact of these mechanisms on behaviour. For each combination of $\alpha$ and behavioural adaptation (no adaptation or mechanisms (A)-(F)), results were observed across ten consecutive runs. For all adaptation mechanisms including reliability-based cue reweighting (C, D, and F), the ten simulation runs from which results were taken, were preceded by runs where the reweighting was allowed to converge from the initial, default sensory weights. It was found that two such simulation runs was enough to ensure convergence. 

A number of additional simulated tests were also run, to investigate the model's sensitivity to the difficulty of the slalom in terms of the distance between cones, the assumed sensory noise levels, and the assumption of intermittent rather than continuous steering control. For this latter test, we replaced the intermittent control part of the model (the right half of \figref{ModelOverview}) with just a fixed time delay of $\Delta T/2 = 0.2$ s and a gain 1/$\Delta T = 2.5$ s$^{-1}$, to obtain a continuous rate of steering wheel change $\dot{\delta}$; as explained in \citep{MarkkulaEtAl2018} this continuous model behaves roughly as an average-filtered version of the intermittent control model. The steering cost for this model was changed, to consider continuous steering wheel rates instead of discrete steering wheel adjustments as in \eqnref{CostFunction}:
\begin{equation}
	\Jsteer = \frac{1}{T}\sum_{i=1}^n \dot{\delta}_i^2,
\end{equation}
and $\ksteer$ was set to 0.1, with the same type of motivation as provided above for the intermittent model.

\section{Results}


\subsection{Impact of motion scaling and behavioural adaptation mechanisms}

\subsubsection{Yaw rate estimation}

\begin{figure}
	\centering
	\includegraphics[width=\columnwidth]{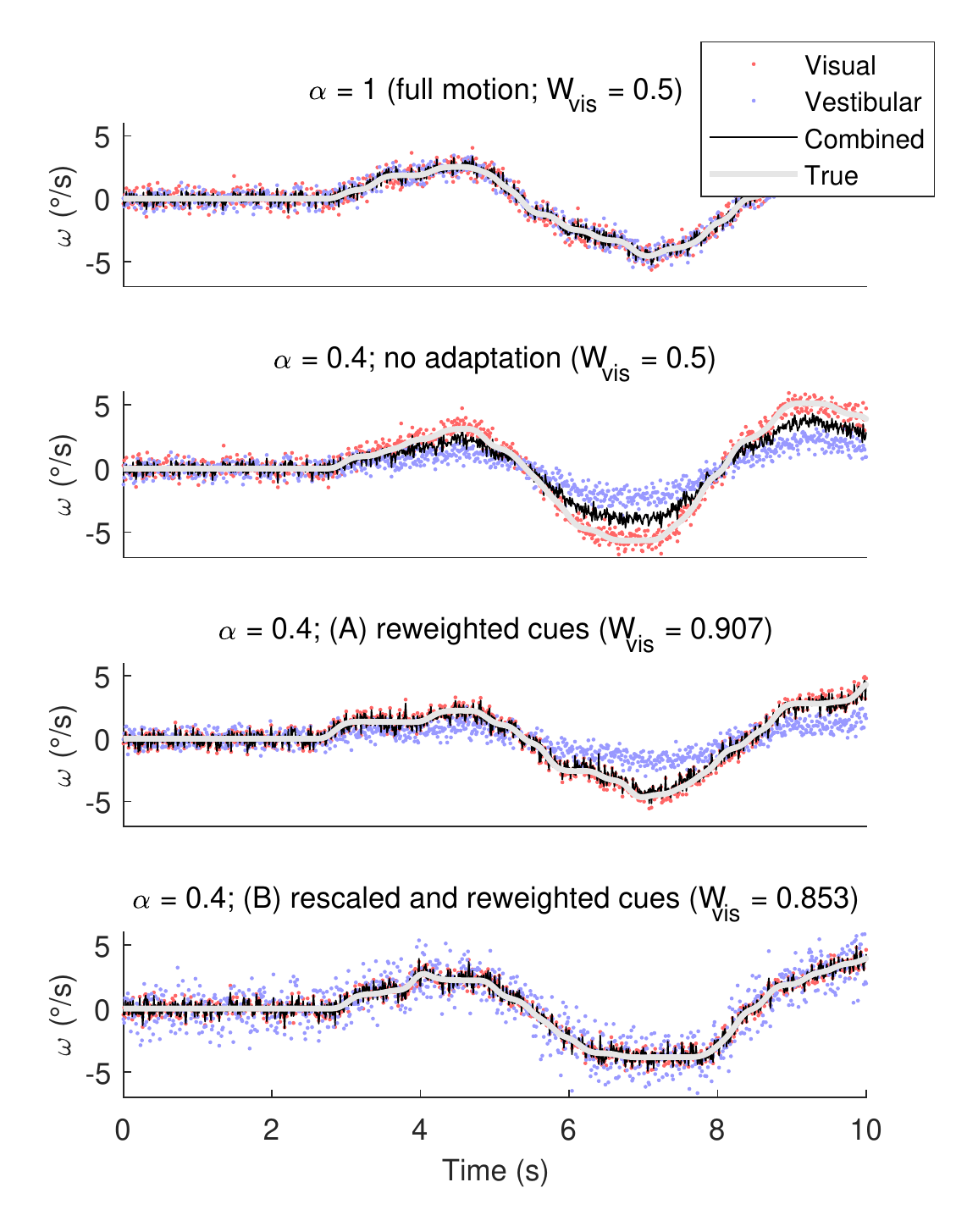}
	\caption{Estimation of yaw rate $\omega$ by the driver model, exemplified by the first ten seconds of four simulations, under different conditions. (A) and (B) refer to two of the adaptation mechanisms described in the text.}
	\label{fig:YawRateEstimation}
\end{figure}

\figref{YawRateEstimation} provides an illustration of the visual-vestibular estimation of yaw rate, from simulations with the model under different conditions. The top panel shows a simulation with full motion cues ($\alpha = 1$), the other three are all simulations with downscaled motion cues ($\alpha = 0.4$). Note that without any adaptation whatsoever (second panel from top), the combined estimate of yaw rate (black line) becomes strongly biased towards zero, away from the true yaw rate (light gray line). This bias all but disappears when reweighting the cues based on prediction-estimated reliabilities (third panel). If rescaling the vestibular cues (bottom panel), a completely non-biased estimate can again be obtained, making what would seem like a theoretically optimal use of the vestibular cue, despite the vestibular noise also becoming visibly inflated by the rescaling. 

\subsubsection{Model time series behaviour}

\begin{figure*}
	\centering
	\includegraphics[width=\textwidth]{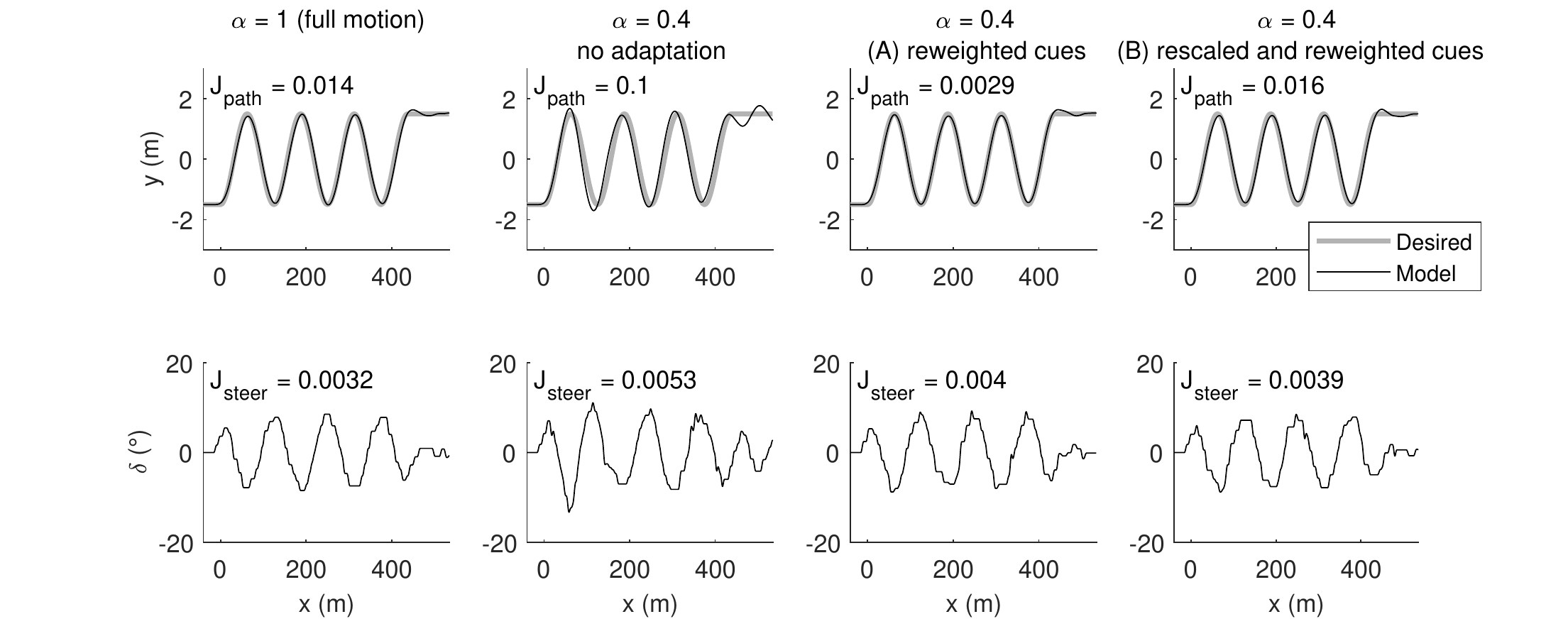}
	\caption{Example time series behaviour of the model in four simulations, under different conditions. The quantities $x$ and $y$ denote, respectively, longitudinal and lateral distance along the slalom track. (A) and (B) refer to two of the adaptation mechanisms described in the text.}
	\label{fig:TimeSeriesBehaviour}
\end{figure*}

\figref{TimeSeriesBehaviour} shows example time series behaviour of the model in the same four conditions as in \figref{YawRateEstimation}. Without any adaptation, the model's steering becomes unstable when motion is scaled down, resulting in increasing path and steering costs. This instability is counteracted when downweighting vestibular cues, both with or without prior rescaling of the vestibular input, but steering efforts still remain higher than in the full motion case. Note, however, that the lowest path cost is obtained for the simulation with reweighted but not rescaled cues.

\subsubsection{Task performance and steering effort}

\begin{figure*}
	\centering
	\includegraphics[width=\textwidth]{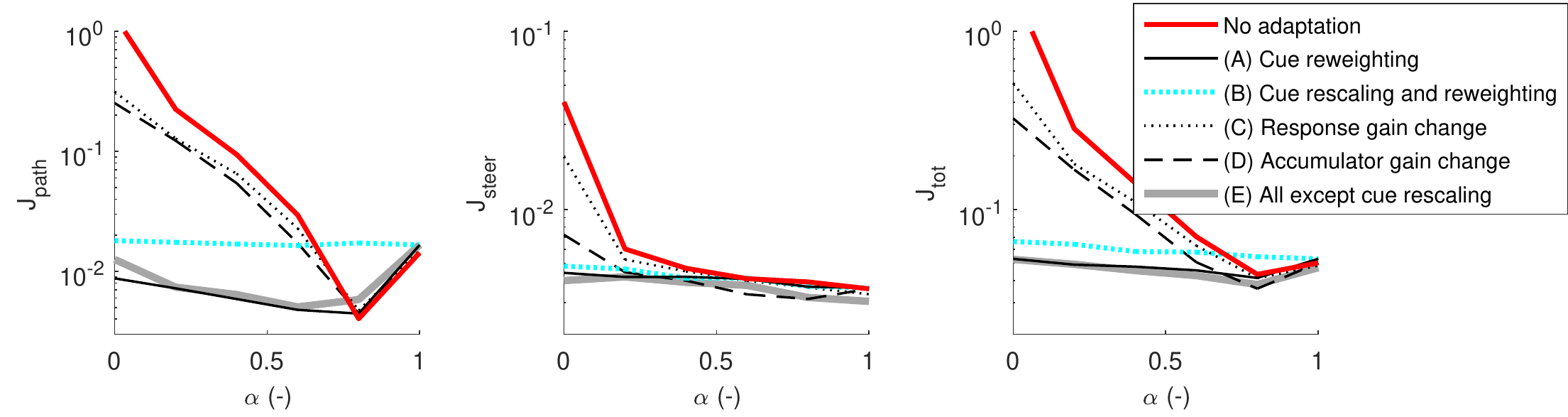}
	\caption{Costs as a function of motion scaling, for the non-adapted model and for the hypothesised adaptation mechanisms (A)-(E).}
	\label{fig:AdaptationMechanismEffects}
\end{figure*}

A more complete overview of path-tracking and steering costs, as a function of motion scaling and adaptation mechanisms, is provided in \figref{AdaptationMechanismEffects}. It can be noted that, in line with the empirical reports, steering efforts ($\Jsteer$) increase with decreasing $\alpha$, especially for the non-adapted model. However, all of the behavioural adaptation mechanisms succeed at improving the situation, which in turn aligns with the empirical reports of decreasing steering efforts after prolonged exposure to the slalom task. 

Related patterns can be observed for the path tracking costs ($\Jpath$), but with the important difference that for all simulations except the cue rescaling adaptation (B), there is a local minimum at $\alpha \in [0.6, 0.8]$, i.e., the model reproduces not only the general tendency of worse performance for down-scaled motion, and improvement with adaptation, but also the sub-unity local optimum for motion scaling that has been reported in the empirical literature. After applying the rescaling adaptation (B), however, there is close to zero effect of motion scaling on the model's task performance.

\begin{figure} 
	\centering
	\includegraphics[width=\columnwidth]{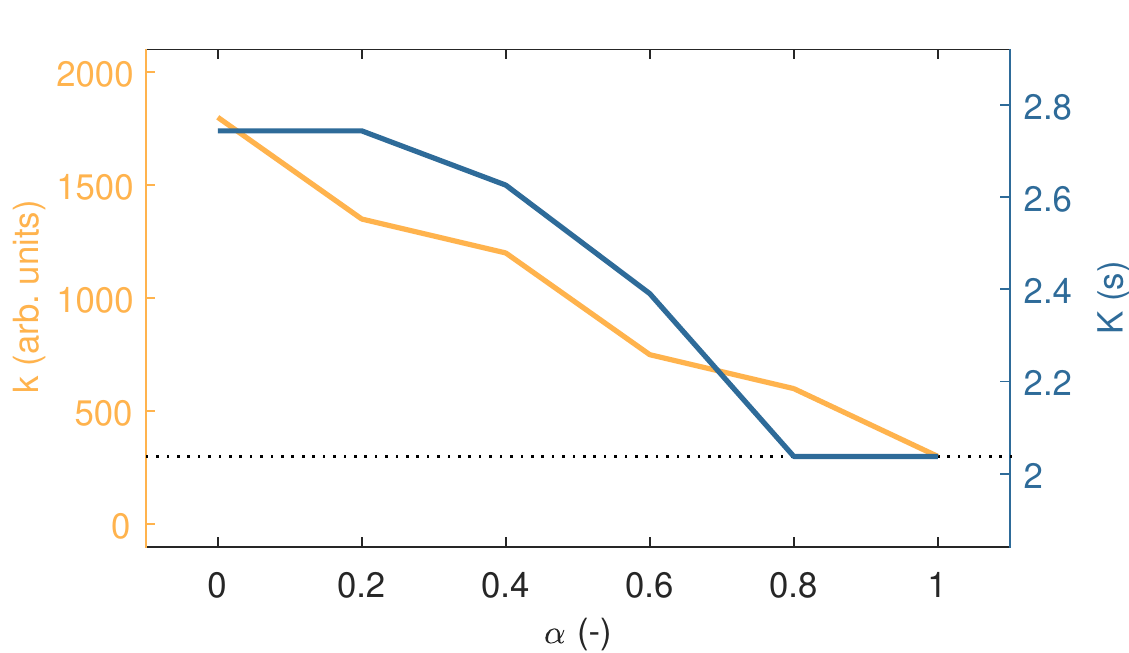}
	\caption{Adapted optimal evidence accumulation gain ($k$) and steering response amplitude gain ($K$), as a function of motion scaling. The dotted horizontal line indicates the optimal gains for full motion ($\alpha = 1$).}
	\label{fig:AdaptedGains}
\end{figure}

\begin{figure}
	\centering
	\includegraphics[width=\columnwidth]{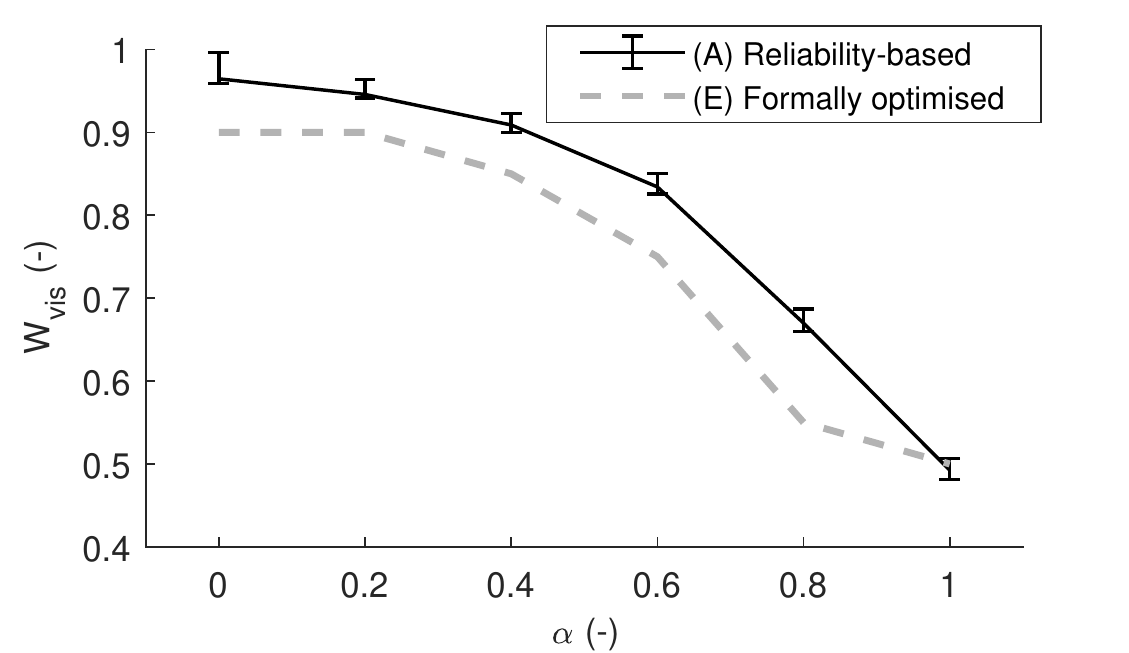}
	\caption{Adapted visual cue weights as a function of motion scaling, estimated from prediction-based sensory reliabilities (shown as average and total spread across ten task repetitions), and by means of formal optimisation. The vestibular weight is not shown, but is always $\Wves = 1 - \Wvis$.}
	\label{fig:AdaptedVisualWeights}
\end{figure}

Analysing the values of adapted parameters provides further insight into how the adaptation mechanisms operate. \figref{AdaptedGains} shows that the performance improvements from adapting evidence accumulation and steering response gains are both had by increasing these gains, resulting in more frequent adjustments and higher-amplitude steering adjustments, respectively. \figref{AdaptedVisualWeights} shows that the cue weighting obtained by measuring reliability as deviations from expected sensory input, a type of reliability estimate that should be readily available to the brain, was relatively close to the optimal weighting obtained when formally optimising for minimal cost $\Jtot$. 

\subsection{Sensitivity to task and model variations}

\subsubsection{Slalom difficulty}

\begin{figure}
	\centering
	\includegraphics[width=\columnwidth]{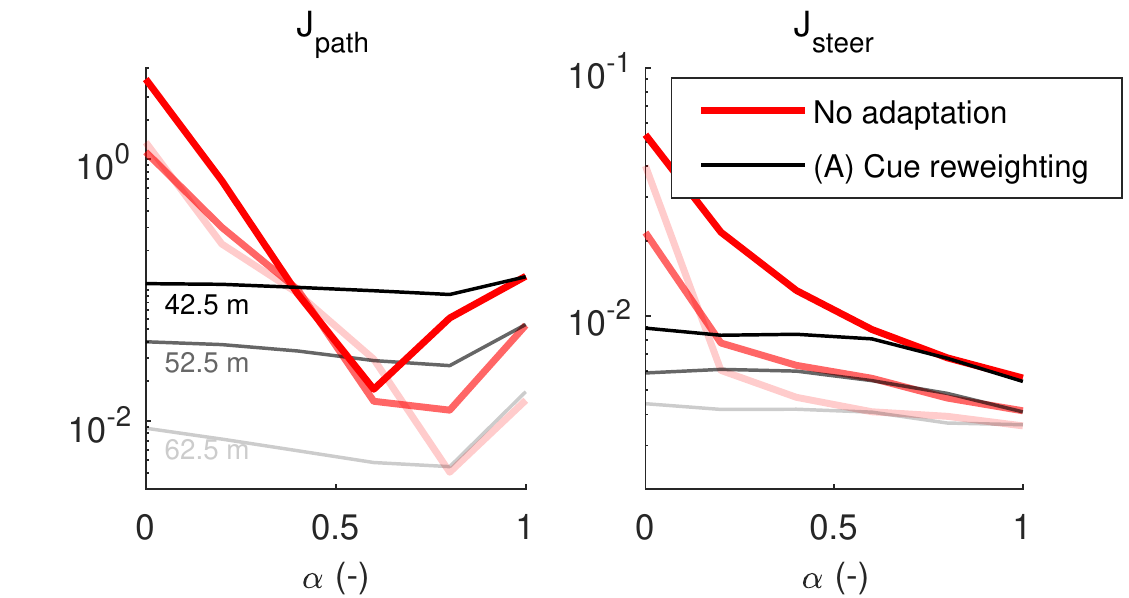}
	\caption{Effect of slalom cone spacing (42.5 m, 52.5 m, or 62.5 m, as indicated in the left panel) on path and steering costs, as a function of motion scaling, without any adaptations and with reliability-reweighted cues.}
	\label{fig:VaryingSlalomDifficulty}
\end{figure}

\figref{VaryingSlalomDifficulty} shows that the qualitative behaviour of the model is insensitive to the spacing of the cones in the slalom task. However, the model can be seen to reproduce another empirical observation \cite{SavonaEtAl2014b}; the motion scaling optimum occurs at lower $\alpha$ for more difficult slaloms (shorter cone spacing, requiring higher lateral accelerations).

\subsubsection{Sensory noise} 

As mentioned earlier, it is difficult to know what magnitudes to choose for the sensory noises. However, as long as $\sigmavis = \sigmaves$, changing these up or down does not change the shape of the path-tracking and steering cost curves; these simply go up and down with the noise levels.

\begin{figure}[t]
	\centering
	\includegraphics[width=\columnwidth]{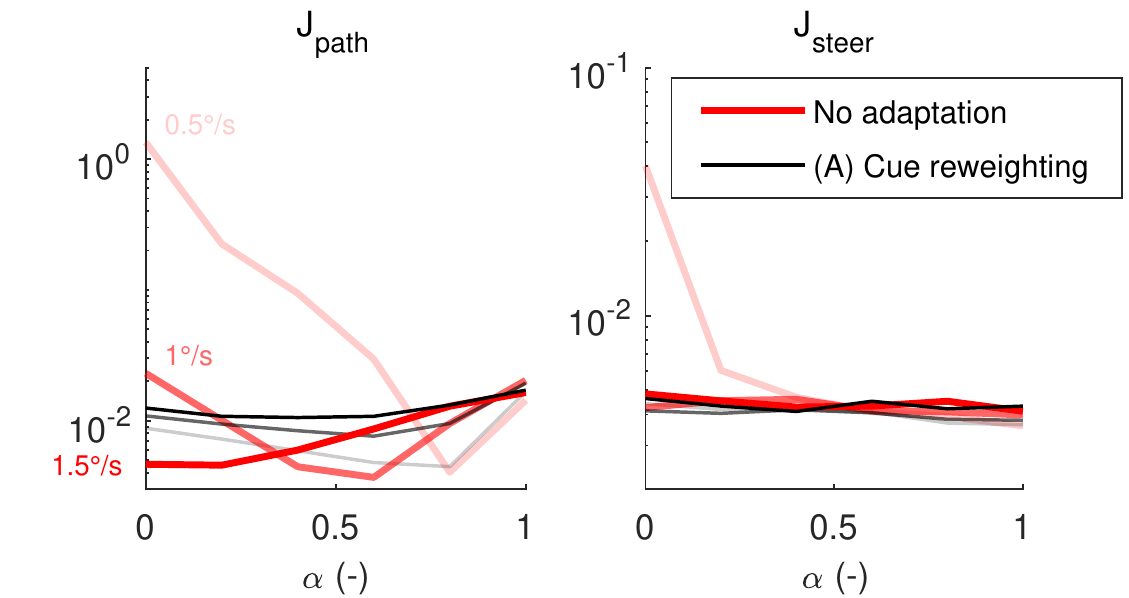}
	\caption{Effect of variations in vestibular sensory noise levels ($\sigmaves$ at 0.5\degree/s, 1\degree/s, or 1.5\degree/s, as indicated in the left panel; $\sigmavis = 0.5\degree$/s throughout) on path and steering costs, as a function of motion scaling, both without any adaptations and with reliability-reweighted sensory cues.}
	\label{fig:VaryingVestibularNoise}
\end{figure}

However, it could be argued, based on perceptual threshold experiments, that the Visual system is more sensitive than the semicircular canals in discrimination of pure yaw motion \cite{Riemersma1981, SoykaEtAl2012}. Therefore, as shown in \figref{VaryingVestibularNoise}, simulations were run where the vestibular noise was increased, while maintaining $\sigmavis = 0.5\degree$/s. The figure shows that doing so maintains the high-level qualitative effects of motion scaling, but the impact of motion scaling is reduced. This is in line with what one may have expected, since with higher vestibular noise levels, \eqnref{OptimalCueIntegration} prescribes lower vestibular sensory weights to begin with, so one needs to reduce $\alpha$ more to see any effects. One consequence of this phenomenon is that the local optimum for path-tracking cost shifts to lower $\alpha$ with increased vestibular noise.

\subsubsection{Continuous model}

\begin{figure}[t]
	\centering
	\includegraphics[width=\columnwidth]{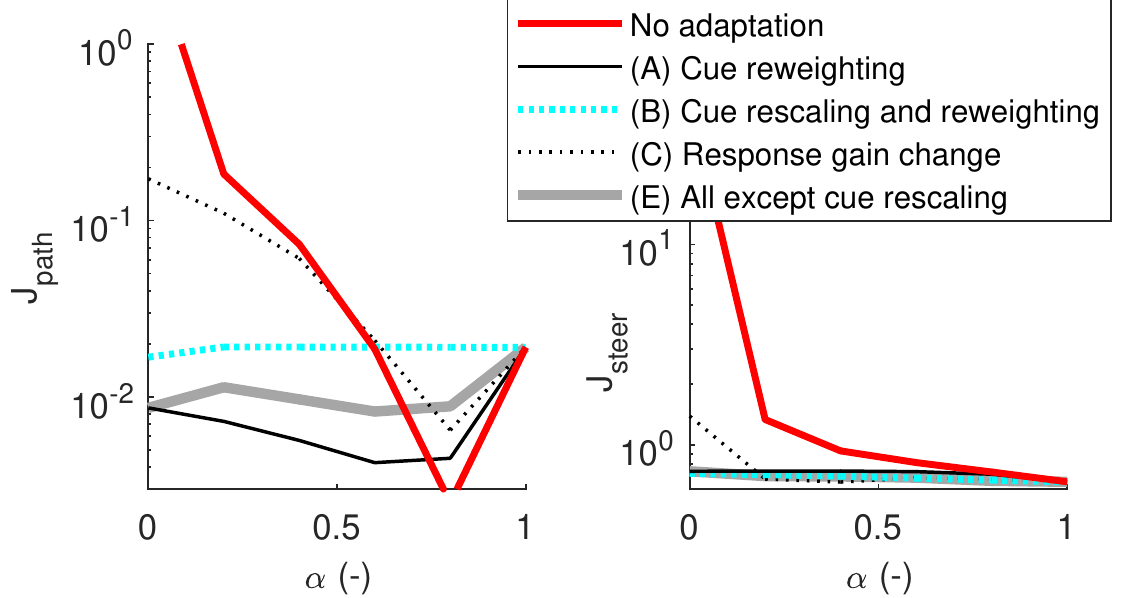}
	\caption{Steering and path costs as a function of motion scaling and adaptation mechanisms, for a simplified, continuous version of the model. The left panel retains the scale of the corresponding panel in \figref{AdaptationMechanismEffects}, but the right panel has a different scale due to the different steering cost function. Adaptation mechanism (D), accumulation gain change, is not applicable to the continuous model.}
	\label{fig:AdaptationMechanismEffects_ContinuousModel}
\end{figure}

Comparing the results for the continuous model in \figref{AdaptationMechanismEffects_ContinuousModel} to those for the intermittent model in \figref{AdaptationMechanismEffects}, it can be seen that many but not all of the qualitative phenomena are retained in the continuous model: Steering costs still go up for the non-adapted model with $\alpha < 1$, and the sub-unity motion scaling optimum for path-tracking persists, but for the continuous model there is close to zero effect of $\alpha$ on steering costs after all of the adaptations.

\section{Discussion}

\subsection{Sensory integration and behavioural adaptation}

The model used here was relatively simple, especially at the level of multisensory integration. Nevertheless, the main targeted empirical phenomena, in terms of task performance and steering efforts, were all qualitatively captured by the model. Thus, more specifically, the model analyses provided here suggest that low task performance and high steering efforts upon first exposure to downscaled motion cues can be understood as drivers responding to these downscaled cues as if they are \emph{directly underestimating vehicle yaw rate}, and use this underestimation when shaping their steering. As a result this steering becomes more effortful, and in many cases also more unstable. 

The fact that the analyses support the idea of drivers using the underestimated yaw rates as part of their control is interesting in its own right. Much of the existing empirical literature on driving simulator motion can be interpreted conservatively in this sense, to say that motion cues mainly cause drivers to change their higher-level strategy in terms of adopted speeds or trajectories, to avoid experiencing large accelerations \citep{SieglerEtAl2001, Jamson2010, CorreiaGracioEtAl2011, BerthozEtAl2013}, or that motion cues mainly support rejection of unexpected external perturbances like wind gusts \citep{RepaEtAl1982, GreenbergEtAl2003}. The present model and simulation analyses, together with the existing empirical findings from slalom experiments, instead suggest a stronger account, whereby drivers make direct use of motion information as part of shaping their steering to reach their intended targets, also in the absence of any external perturbances or high-level adaptations of trajectory or speed. It is clear that existing multisensory models of drivers \citep{NashAndCole2018} and pilots \cite[e.g., ][]{MulderEtAl2013} also suggest this deep form of involvement of vestibular cues in control, but we are not aware of any prior analysis of empirical data providing support for it in driving.

With respect to the behavioural adaptation, the results presented here indicate that especially three mechanisms, or several of them in combination, are candidates for causing the empirically observed effects of repeated task exposure: increased gains in evidence accumulation or steering response, or sensory cue reweighting based on reliabilities inferred from deviations between received and predicted sensory input. Especially this latter mechanism seems readily implementable in neural systems, and it also provided the most dramatic performance improvements. Interestingly, the cue weights obtained in this way came close to the formally optimal values (\figref{AdaptedVisualWeights}); that this would be the case was not clear a priori, since the formally optimal values depend on the specific task and chosen cost function.

Also the increases in accumulator and steering gains can actually be interpreted in neurobiological terms; increases in global cortical arousal by means of broad diffusion of noradrenaline (also known as norepinephrine) has been found to have the specific effect of increasing gains in neuronal response to inputs \citep{AstonJonesAndCohen2005}. Thus, it coulde be hypothesised that the driver's brain may respond to unsatisfactory task performance with release of noradrenaline, increasing the evidence accumulation and response gains.

The present analyses do not provide any conclusive insight into which of the three adaptation mechanisms mentioned above may have been more  important in the empirical studies. However, the results can possibly be taken to suggest that complete rescaling of the vestibular input may not have occurred in these studies, since if so there should not, according to the model simulations, have remained any local optimum in task performance for $\alpha < 1$. This leads on to the next section.

\subsection{Sub-unity optimum in motion scaling}

It was not expected beforehand that the model would exhibit the local performance optimum for sub-unity motion scaling. Previously, it has been proposed that this phenomenon might be due to imperfections and false cues in motion systems becoming more prominent at $\alpha$ too close to 1, or to motion downscaling being needed to ensure coherence with underestimated visual speeds \cite{BerthozEtAl2013}, or to successful control of arms and steering wheel becoming more difficult when the body is subjected to higher and more uncomfortable accelerations \cite{SavonaEtAl2014b}. However, none of these mechanisms were included in the model, yet it still reproduced the phenomenon.

Here, what is happening instead is that the model gets a path-tracking benefit from slightly underestimating the yaw rate in this task. Looking closely at the different vehicle trajectories in \figref{TimeSeriesBehaviour} it can be seen that the path tracking cost is notably affected by how well the vehicle's path is phase-aligned with the desired path; e.g., in the full motion example (leftmost panel) the vehicle trajectory reaches apex slightly after the desired path, whereas in the downscaled and reweighted cues example (third panel from left) the apexes align almost perfectly, resulting in a considerably lower path tracking cost. Interestingly, when \citet{BerthozEtAl2013} analysed what caused the path tracking optimum for human drivers in their study, this type of trajectory phasing is exactly what they found, also with increasingly early vehicle path apexes for increasingly downscaled motion cues. In the model, the reason this is happening is that the more the model underestimates its own yaw rate, the earlier and the more vigorously it will respond to the next turn in the slalom, causing a shift toward earlier phase. This explanation also aligns with the analysis of steering angles by \citet{FeenstraEtAl2010}; their observed faster adjustment of steering just as passing the cone when driving without motion is indeed what one would expect from a driver who perceives that s/he isn't rotating quickly enough in preparation for the next cone. 

Overall, the fact that the model captures so many aspects of the sub-unity optimum phenomenon--not only the existence of an optimum, but also its cause being trajectory phasing, and in addition also the effect of decreasing optimum $\alpha$ for more difficult slaloms (\figref{VaryingSlalomDifficulty})--seems to suggest that underestimation of yaw rate (or, as said before, behaving as if one underestimates yaw rate) may indeed be the mechanism behind this phenomenon also in humans. 

However, it should be noted that the local optimum in motion scaling has been observed also in subjective ratings of various types, and not necessarily at the exact same $\alpha$ as the path-tracking optimum \citep{BerthozEtAl2013, SavonaEtAl2014b}. While the present model simulations can \emph{possibly} help explain why there might be a local optimum in subjective ratings of for example \inquotes{ease of completing the task} or similar, it seems less clear how the yaw rate underestimation mechanism would contribute to improved subjective ratings of for example realism. One possibility is of course that several mechanisms are in play here; the yaw rate underestimation mechanism might coexist with any and all of the mechanisms reviewed above in this section.

Another question raised by the present analyses is to what extent the sub-unity motion scaling optimum phenomenon is particular to slalom tasks.  The path-tracking benefit from yaw rate underestimation observed here seems to be rather closely related to the specific nature of this task, such that it might not necessarily generalise.

\subsection{Driver steering modelling}

We have shown in this paper how the intermittent control framework of \citet{MarkkulaEtAl2018} can be naturally extended to account for multisensory integration and a range of behavioural adaptation mechanisms; these extensions should be of value in a range of contexts relating to driving simulators and in other application domains. 

It is notable, however, that much of the results of the intermittent control model were reproduced also by the continuous model. The main differences, quite expectedly, related to steering effort: In addition to the lack of the putatively effort-related evidence accumulation gain adaptation, which cannot be included in the continuous model, the effect of motion scaling on steering costs all but disappeared with all behavioural adaptations in the continuous model. Nevertheless, since the continuous model is so simple and easy to implement (just the few operations in the left part of \figref{ModelOverview} plus a delay and a gain), it seems particularly well suited for use in future studies of multisensory integration in driving simulators.

The main alternative to the models tested here is the one proposed by \citet{NashAndCole2018}. This model assumes optimal control in a more complete sense, with access to explicit internal models of the vehicle as well as of the sensory and neuromuscular dynamics. On a cursory analysis, it seems likely that also this model would exhibit lower performance and higher steering effort when motion is scaled down or turned off. Among the adaptation mechanisms studied here, the combined rescaling and reweighting mechanism (B) would seem to fit especially naturally within the framework. Here, the model behaviour with this mechanism did not align well with the existing empirical findings, but direct testing in model simulation would be needed to see whether it fares better in the \citet{NashAndCole2018} framework. Also the pure cue reweighting adaptation (A) could be used, if the optimal control framework is extended with a similar prediction-based estimation of reliabilities as proposed here. However, the gain adaptations (C) and (D) do not seem readily compatible with the \citet{NashAndCole2018} model.

It would also be interesting to test whether the \citet{NashAndCole2018} model reproduces the sub-unity motion scaling optimum phenomenon. At least in its non-adapted form, it should also underestimate yaw rate, but it might be that the very reason our model tracks the desired path better when underestimating yaw rate is because even with full motion cues its control is to some degree inherently suboptimal, in contrast with the \citet{NashAndCole2018} model.

\subsection{Future work}

As has already been hinted at above, one obvious direction would be to apply the models proposed here to other tasks besides the slalom. Ideally, one would first generate predictions for these other tasks using the model (perhaps just the continuous version of it, for simplicity), and then investigate whether these predictions are borne out in tests with human drivers. In such empirical work, one could also study the adaptation process itself in more detail, for example to try to distinguish better between the various hypothesised mechanisms for adaptation.

The model proposed here could also be applied in development and tuning of motion cueing algorithms, in its present form especially for direct scaling cueing. For example, one could use the model to predict what motion scaling might yield behaviour that is as similar as possible to that in a real vehicle, or to predict how large the differences might be between behaviour before and after adaptation to a certain motion scaling, to get an idea of what motion cueing settings might take longer time to get used to. 

If the model is to be used with more complex, arbitrary motion cueing algorithms, for example affecting rotations and translations in different ways, the current simplified approach of capturing all vestibular sensing in just a yaw rate estimate will not be enough, and it needs to be better considered how different types of motion cues are used by drivers when determining the needed steering control. One possibility is the \citet{NashAndCole2018} approach, with sensory dynamics models and inverse model state estimators (modelling scheme (c) in \figref{MultisensoryIntegrationSchemes}). A possible complication here is that empirical observations suggest that rotation and translation cues may not be used by human drivers in precisely the ways suggested by this type of engineering analysis \citep{LakerveldEtAl2016}.

\section{Conclusion}

We aimed to study visual-vestibular integration and behavioural adaptation in the driving simulator, by modelling human steering behaviour in slalom driving, a type of task that has been much studied in simulators with linearly downscaled motion cues. To this end, we extended on an existing framework for intermittent sensorimotor control; this extended framework, as well as a simplified continuous version of it, seem well suited for future work on multisensory integration and behavioural adaptation in simulators and elsewhere.

A key insight from the present work is that much of the empirically observed phenomena in motion downscaling of slalom tasks can be explained by a \emph{yaw rate underestimation} mechanism, whereby drivers respond to downscaled simulator motion as if they perceived themselves and the simulated vehicle to be rotating less than is the case in the simulated situation. This mechanism explains not only why removing motion cues leads to increased steering efforts and worse path tracking, but also captures a phenomenon for which there has previously been no good explanation: In an intermediate range of mild downscaling, the yaw rate underestimation phenomenon counteracts an inherent suboptimality in the model's steering behaviour, resulting in improved path tracking.

Empirically observed effects of task exposure have also been studied, by means of a range of hypothesised and neurobiologically plausible behavioural adaptation mechanisms. Out of these mechanisms, the only one that did not provide behavioural changes in line with the existing literature was a compensatory rescaling of the vestibular input (to counteract the motion downscaling). The results instead suggest that empirically observed improvements over time, in prolonged task performance with down-scaled cues, may be achieved by drivers by increasing gains in sensory evidence accumulation or steering response, and/or by detecting deviations between actual vestibular input and internal predictions of the same, and then using these deviations to downweight vestibular cues in an optimal cue integration scheme.

\section*{Acknowledgments and data statement}

This work was supported by Jaguar Land Rover and the UK-EPSRC Grant EP/K014145/1 as part of the jointly funded Programme for Simulation Innovation (PSi). There is no primary research data associated with this paper; as should be clear from the text, all of the comparisons to human behaviour refer to empirical findings reported in previous literature.


		%
%




\bibliographystyle{elsarticle-harv} 
\bibliography{Bibliography_DSC2018Multisensory}

\end{document}